\documentclass[10pt]{article}

\usepackage{amsmath}
\usepackage{amssymb}
\usepackage[english]{babel}
\usepackage{graphicx}

\usepackage{cite}

\usepackage{color}
\usepackage{caption}
\usepackage{subcaption}

\makeatletter
\renewcommand{\@biblabel}[1]{\quad#1.}
\makeatother

\date{}

\begin{document}

\begin{flushleft}
{\Large \textbf{Illumina Sequencing Artifacts Revealed by Connectivity
    Analysis of Metagenomic Datasets} }
\\
Adina Chuang Howe$^{1,2}$, 
Jason Pell$^{3}$,
Rosangela Canino-Koning$^{3}$,
Rachel Mackelprang$^{4}$,
Susannah Tringe$^{4}$,
Janet Jansson$^{4,5}$ ,
James M. Tiedje$^{1,2}$, and 
C. Titus Brown$^{1,3\ast}$
\\
\bf{1} Microbiology and Molecular Genetics, Michigan State University, East Lansing, MI, USA
\\
\bf{2} Plant, Soil, and Microbial Sciences, Michigan State University, East Lansing, MI, USA
\\
\bf{3} Computer Science and Engineering, Michigan State University, East Lansing, MI, USA
\\
\bf{4} Department of Energy (DOE) Joint Genome Institute, Walnut Creek, CA, USA
\\
\bf{5} Lawrence Berkeley National Laboratory, Genomics Division, Berkeley, CA, USA
\\
$\ast$ E-mail: ctb@msu.edu
\end{flushleft}

\section*{Abstract}

Sequencing errors and biases in metagenomic datasets affect
coverage-based assemblies and are often ignored during analysis.
Here, we analyze read connectivity in metagenomes and identify the
presence of problematic and likely a-biological connectivity within
metagenome assembly graphs.  Specifically, we identify highly
connected sequences which join a large proportion of reads within each
real metagenome.  These sequences show position-specific bias in
shotgun reads, suggestive of sequencing artifacts, and are only
minimally incorporated into contigs by assembly.  The removal of these
sequences prior to assembly results in similar assembly content for
most metagenomes and enables the use of graph partitioning to
decrease assembly memory and time requirements.

\section*{Introduction}

With the rapid decrease in the costs of sequencing, we can now achieve
the sequencing depth necessary to study microbes from even the most
complex environments \cite{Hess:2011p686,Qin:2010p189}.  Deep
metagenomic sequencing efforts in permafrost soil, human gut, cow
rumen, and surface water have provided insights into the genetic and
biochemical diversity of environmental microbial populations
\cite{Hess:2011p686,Iverson:2012p1281,Qin:2010p189} and their involvement in responding to environmental changes
\cite{Mackelprang:2011p1087}. These metagenomic studies have all
leveraged \emph{de novo} metagenomic assembly of short reads for
functional and phylogenetic analyses. \emph{De novo} assembly is an
advantageous approach to sequence analysis as it reduces the dataset
size by collapsing the more numerous short reads into fewer contigs
and enables improved annotation-based approaches by providing longer
sequences \cite{Miller:2010p226,Pop:2009p798}. Furthermore, it does
not rely on the {\em a priori} availability of reference genomes to enable
identification of gene content or operon structure
\cite{Hess:2011p686,Iverson:2012p1281}.

Although \emph{de novo} metagenomic assembly is a promising approach
for metagenomic sequence analysis, it is complicated by the variable
coverage of sequencing reads from mixed populations in the environment
and their associated sequencing errors and biases
\cite{Mende:2012p1262,Pignatelli:2011p742}. Several
metagenome-specific assemblers have been developed to deal with
variable coverage communities, including Meta-IDBA
\cite{Peng:2011p898}, MetaVelvet \cite{Namiki:2012iq}, and SOAPdenovo \cite{Li:2010p234}.  These
assemblers rely on analysis of local sequencing coverage to help build
assemblies and thus are sensitive to the effects of sequencing errors
and biases on coverage estimations of the underlying dataset. The
effects of sequencing errors on \emph{de novo} assembly has been
demonstrated in simulated metagenomes
\cite{Mavromatis:2006p894,Mende:2012p1262,Pignatelli:2011p742} and
isolate genomes \cite{Morgan:2010p740,Chitsaz:2011kr}, but these datasets do not necessarily represent real metagenomic
data. Specifically, these models exclude the presence of known
non-biological sequencing biases which hinder assembly approaches
\cite{GomezAlvarez:2009p1334,Keegan:2012p1336,Niu:2010p1333}.

In this study, we examine metagenomic datasets for the presence of
artificial sequencing biases that affect assembly graph structure,
extending previous work to large and complex datasets produced from
the Illumina platform. We characterize sequence connectivity in an
assembly graph, identifying potential sequencing biases in regions
where numerous reads are connected together.  Within metagenomic
datasets, we find that there exist highly connected sequences which
partially originate from sequencing artifacts.  Moreover, these
sequences limit approaches to divide or partition large datasets for
further analysis, and may introduce artifacts into assemblies.  Here,
we identify and characterize these highly connected sequences and
examine the effects of removing these sequences on downstream
assemblies.

\section*{Results}

\subsection*{Connectivity analysis of metagenome datasets}

\subsubsection*{Presence of a single, highly connected lump in all datasets}
We selected datasets from three medium to high diversity
metagenomes from the human gut \cite{Qin:2010p189}, cow rumen
\cite{Hess:2011p686}, and agricultural soil (SRX099904 and SRX099905)
(Table~\ref{data-summary}).  To
evaluate the effects of sequencing coverage, we included two subsets
of the 520 million read soil metagenome containing 50 and 100 million
reads.  We also included a previously published error-free simulated
metagenome based on a mixture of 112 reference genomes
\cite{Pignatelli:2011p742}.

We evaluated read connectivity by partitioning reads into disconnected
components with a de Bruijn graph representation \cite{Pell:2012cq}.  This approach
guarantees that reads in different partitions do not connect to each
other and permits the separate assembly and analysis of each
partition.  For each metagenome, regardless of origin, we found a
single dominant, highly connected set of sequencing reads which we
henceforth refer to as the ``lump'' of the dataset (Table~\ref{data-summary}).  This lump contained the largest subset of connected sequencing
reads and varied in size among the datasets, ranging from 5\% of total
reads in the simulated metagenome to 75\% of total reads in the human
gut metagenome.  For the soil datasets, as sequencing coverage (e.g.,
the fraction of reads mapped to an assembly) increased from 1.4 to 4.7
to 5.6\%, the lump size increased more dramatically from 7 to 15 to
35\% of all reads, indicating increasingly larger connectivity between
sequences with more sequencing.

\subsubsection*{Characterizing connectivity in the dominant partition}

We characterized the connectivity of sequences
within each lump by estimating the average local graph density from
each k-mer (k=32 unless otherwise stated) in the assembly graph (see
Methods).  Here, local graph density is a measurement of total
connected reads within a fixed radius.  Sequences
in the identified metagenomic lumps were characterized by very high
local graph densities: between 22 to 50\% of the total nodes in
metagenomic lump assembly graphs had average graph densities greater
than 20 (Table~\ref{data-summary}).  This indicates that these nodes were in very nonlinear portions of the assembly graph and had high connectivity.  In comparison, 17\% of the total nodes in the
simulated lump had an average local graph density greater than 20, and
fewer than 2\% of the nodes in the entire simulated data (all partitions) set had an
average graph density higher than 20.

\begin{figure}[h]
\center
{\includegraphics[width=5in]{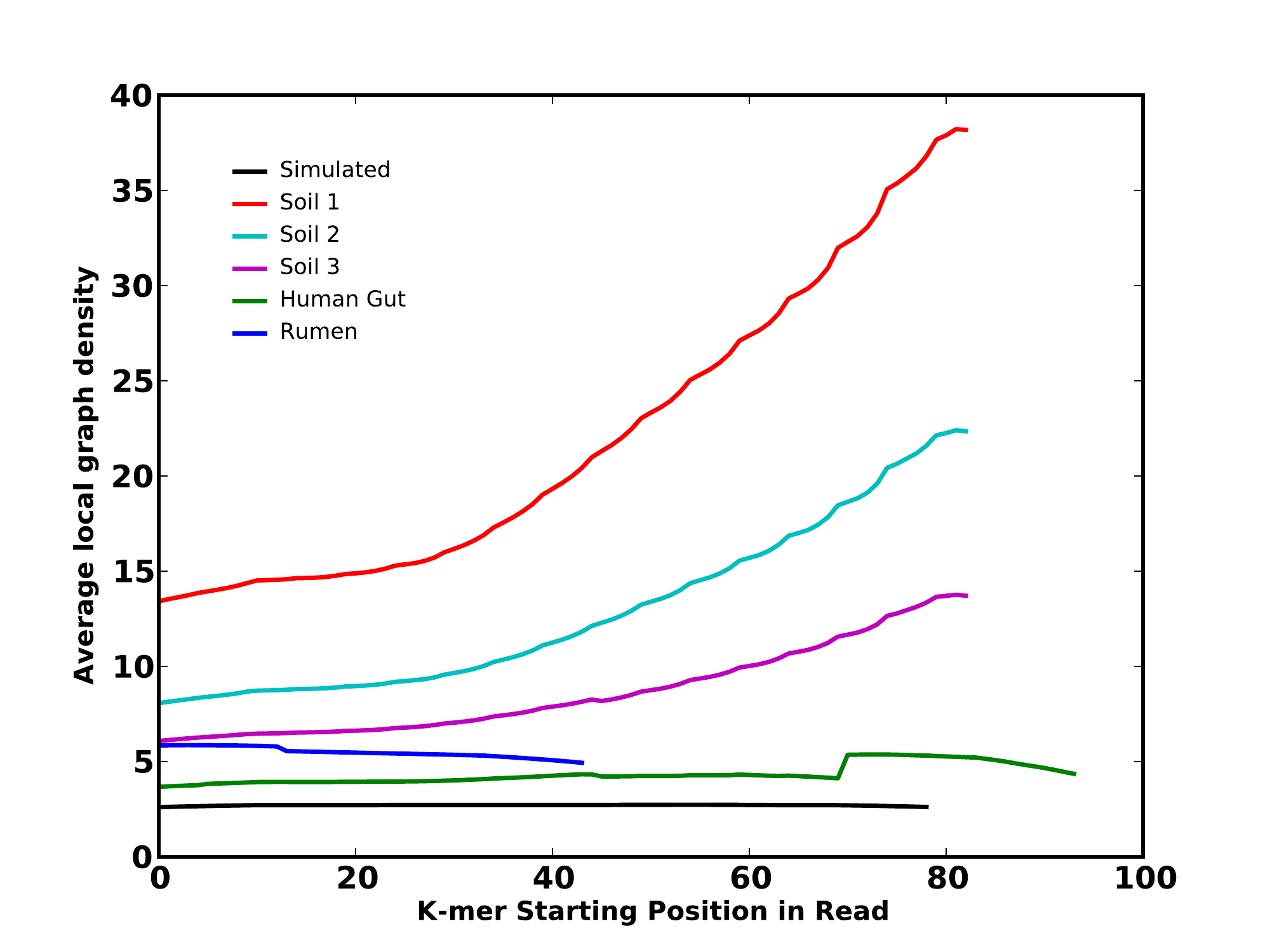}}
\caption{The extent to which average local graph density varies by read position is shown for the lump of various datasets.}
\label{density-pos}
\end{figure}

We next assessed the extent to which graph density varied by position
along the sequencing reads.  The degree of position-specific variation of
graph densities was estimated by calculating the average local graph
density within ten steps of every k-mer by position in each read.  In
all environmental metagenomic reads, we observed variation in graph
density at the 3'-end region of reads (Fig.~\ref{density-pos}).  In soil
metagenomes, we observed the most dramatic variation with local graph
density increasing in sequences located at the 3'-end of the reads.
Notably, this trend was not present in the simulated dataset.

Next, we performed an exhaustive traversal of the assembly graph and
identified the specific sequences within dense regions of the assembly
graph which consistently contributed to high connectivity.  We
observed that this subset of sequences was also found to exhibit
position-specific variation within sequencing reads, with the
exception of these sequences in the simulated dataset (Fig.~\ref{pos-spec}, solid
lines).  As with local density trends, position-specific trends in
the location of these sequences also varied between metagenomes.  As
sequencing coverage increased among metagenomes, the amount of 3'-end
variation appeared to decrease (e.g., the soils) or increase (e.g.,
rumen and human gut).

\subsection*{Effects of removing highly connected sequences on assembly}

\subsubsection*{Removal of highly connected sequences enables graph partitioning of metagenome}

Since these highly connected sequences exhibited position-specific
variation indicative of sequences of non-biological origin, we removed
them and assessed the effect of their removal on assembly (see
Methods).  We found that by removing these k-mers, we could
effectively break apart metagenomic lumps, and the resulting largest
partition of connected reads in each metagenome was reduced to less
than 7\% of the total reads in the lump.  Partitioning also had the
effect of significantly decreasing assembly time and memory usage
\cite{Pell:2012cq}.

\subsubsection*{Removing highly connected sequences resulted in minimal losses of reference genes}

We explored the extent
to which the identified highly connected
sequences impacted assembly by first evaluating the effects of the
removal of these sequences from the simulated lump.  The assembly of the reads in the original,
unfiltered simulated lump and that of the reads remaining after
removing highly connected sequences (the filtered assembly) were
compared for three assemblers: Velvet \cite{Zerbino:2008p665}, Meta-IDBA \cite{Peng:2011p898}, and SOAPdenovo \cite{Li:2010p234}.
Based on the total assembly length of contigs greater than 300 bp,
filtered assemblies of the simulated metagenome resulted in a loss of
between 4 - 16\% of total assembly length (Table~\ref{assembly-stats}).  In general, the
filtered assemblies contained fewer total contigs than unfiltered
assemblies, while the maximum contig size increased in the Velvet
assembly but decreased in the Meta-IDBA and SOAPdenovo assemblies.
Direct comparisons of the unfiltered and filtered simulated metagenome assemblies found
that the filtered assemblies comprised on average 89\% of the
unfiltered assemblies, and the unfiltered assemblies contained nearly
all (97\%) of the filtered assembled sequences.  Despite the removal
of over 3\% of the total unique 32-mers in the simulated metagenome,
the resulting filtered assemblies lost only 3-15\% of annotated original reference genes (Table~\ref{assembly-compare}).

We next evaluated the effects of removing highly connected sequences in
real datasets.  Similar to the simulated assemblies, the
removal of highly connected sequences for all metagenomes and
assemblers resulted in a decrease of total number of contigs and assembly
length (Table~\ref{assembly-stats}).  In general, filtered assemblies were largely
contained within unfiltered assemblies and comprised 51-87\% of the
unfiltered assembly.  The observed changes in metagenomic assemblies
were difficult to evaluate as no reference genomes exist, 
and a decrease in assembly length may actually be beneficial if it
eliminates contigs that incorporate sequencing artifacts.
To aid in this evaluation, we used the previously published set of
rumen draft genomes from \emph{de novo} assembly efforts of high
abundance sequences in the rumen metagenome \cite{Hess:2011p686}.
Overall, we found that removal of highly connected sequences from the
rumen dataset resulted in 9-13\% loss of sequences present in
draft reference genomes (Table~\ref{assembly-compare}).

\subsubsection*{Unfiltered assemblies contained only a small fraction of highly connected sequences}

\begin{figure}[h]
\center{\includegraphics[width=\textwidth,height=\textheight,keepaspectratio]{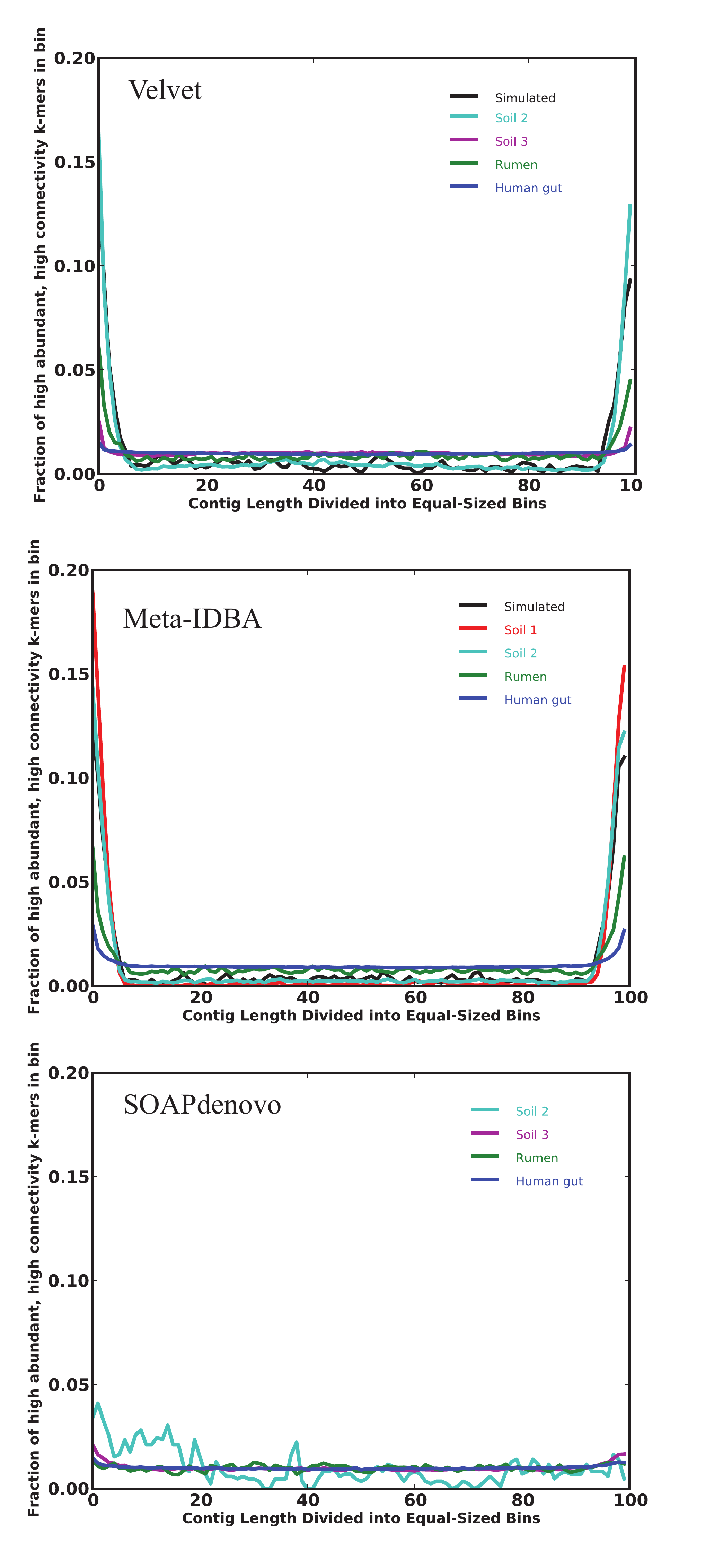}}
\caption{When incorporated into an assembly, abundant (greater than 50 times), highly connecting sequences (k-mers) were disproportionately present at the ends of contigs.  The total fraction of highly connecting k-mers which are incorporated into each contig binned region.}
\label{stoptag-contig}
\end{figure}

To further study the effects of highly connected sequences, we
examined their incorporation into unfiltered assemblies.  Except in
the human gut sample, fewer than 2\% of highly connected sequences
were incorporated by any assembler (Table~\ref{assembly-stoptags}).  Each assembled
contig was divided into percentile bins and examined for the
presence of the previously identified highly connected sequences.  We
found that contigs, especially in assemblies from Velvet and
Meta-IDBA, incorporated a larger fraction of these sequences at their
ends relative to other positions (Fig~\ref{stoptag-contig}).  The SOAPdenovo
assembler incorporated fewer of the highly connected sequences into
its assembled contigs; in the simulated data set, none of these sequences
were assembled, and in the small soil data set only 41 were assembled.  For
the human gut metagenome assemblies, millions of the highly connected
sequences were incorporated into assembled contigs, comprising nearly
4\% of all assembled sequences on Velvet contig ends (Fig~\ref{stoptag-contig}).

\subsubsection*{Identifying origins of highly connected sequences in known reference databases}

For the simulated metagenome, we could identify the source of highly
connected k-mers using available reference genomes. Reference genes
with multiple perfect alignments to highly connected k-mers present in
the dataset a minimum of 50 times were identified (Table~\ref{sim-stoptags}).  Many of
these sequences were from well-conserved housekeeping genes involved
in protein synthesis, cell transport, and signaling.  To determine
possible biological sources of highly connected sequences within real
metagenomes, we compared the sequences shared between the soil, rumen,
and human gut metagenomes (a total of 241 million 32-mers).  Among these 7,586 shared sequences, we identified the closest reference
protein from the NCBI-nr database requiring complete sequence
identity.  Only 1,018 sequences (13\%) matched existing reference
proteins, and many of the annotated sequences matched to
genes conserved across multiple genomes.  The most abundant
proteins conserved in greater than 3 genomes are shown in (Table~\ref{meta-stoptags}), and
largely encode for genes involved in protein biosynthesis, DNA
metabolism, and biochemical cofactors.

One potential cause of artificial high connectivity within metagenomes
is the presence of high abundance subsequences.  Thus, we identified the
subset of highly connected k-mers which were also present with an
abundance of greater than 50 within each metagenome and their location
in sequencing reads (Fig~\ref{pos-spec}, dotted lines).  These high abundance
k-mers comprised a very small proportion of the identified highly
connected sequences, less than 1\% in the soils, 1.5\% in the rumen,
and 6.4\% in the human gut metagenomes, but the position-specific
variation of these sequences was very similar to the variation in the
larger set of highly connected k-mers.

\begin{figure}[h]
\center{\includegraphics[width=\textwidth,height=\textheight,keepaspectratio]{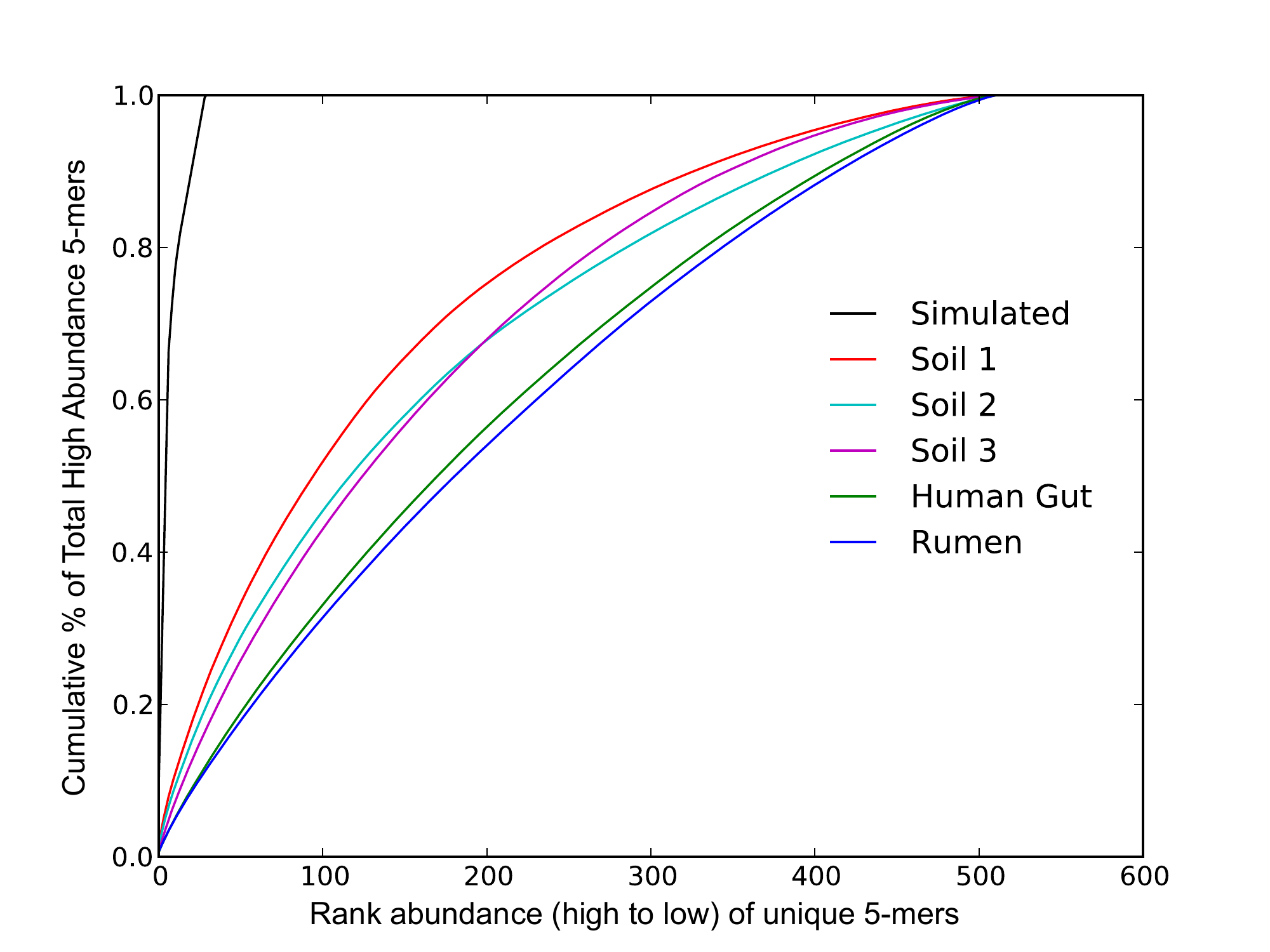}}
\caption{Rank abundance plot of 5-mers present in abundant, highly connected sequences in various datasets.}
\label{five-mer}
\end{figure}

We attempted to identify patterns within the sequences causing
position specific variation by examining the abundance distribution of
5-mers within the highly abundant/highly connected 32-mers.  There
were significantly fewer 5-mers in the simulated sequences compared to
those in metagenomes: 336 total 5-mers in the simulated data and from
425,572 to 221,085,228 total 5-mers in the small soil and human gut
datasets, respectively.  In the simulated dataset, the top ten most
abundant unique k-mers made up 75\% of the total 5-mers; in contrast,
in the metagenomes, k-mers were more evenly distributed: the top ten
most abundant 5-mers comprised less than 10\% of the total 5-mers.
The cumulative abundance distribution of the ranked 5-mers shown in
Fig.~\ref{five-mer} shows this even distribution in all of the real metagenomes.
This suggests that there is no single, easily-identifiable set of
sequences at the root of the highly connected component observed in
real metagenomes.

\section*{Discussion}

\subsection*{Sequencing artifacts are present in real metagenomes}

Through assessing the connectivity of reads in several metagenomes, we
identified a disproportionately large subset of reads
connected together within an assembly graph, which we refer to as
the ``lump.''
The total number of reads in
metagenomic lumps (7-75\% of reads) was significantly larger than that
of simulated dataset (5\% of reads) (Table~\ref{data-summary}).  In the simulated data,
this component consists
of reads connected by
sequences conserved between multiple genomes
(Table~\ref{sim-stoptags}). The larger size of this component
within the soil, rumen, and human gut metagenomes
suggests that anomalous, non-biological connectivity may be present
within these lumps.  Moreover, in the soil metagenomes, we
observed that for a 5\% increase in sequencing coverage of the assembled
contigs, the amount of connectivity nearly doubled.
While sequencing coverage of assembled contigs
increased slightly from 4.7 to 5.6\% in the medium and large soil
metagenomes, the number of reads contained in the lump grew
significantly from 15 million to 182 million.  Given the very high
diversity and very low coverage of these soil samples, the magnitude of the
observed increases in connectivity cannot be due simply to increased
coverage.  This suggests the presence of non-biological features
that falsely connect reads.

The superlinear increase in connectivity exhibited in these data sets
indicates that a form of preferential attachment is occurring in the
graphs \cite{Barabasi:1999p1083}.  This graph-theoretic phenomenon
describes the results of a process where highly connected nodes in a
graph preferentially acquire new edges -- colloquially known as ``the
rich get richer.''  In assembly, any systematic bias towards producing
specific subsequences from shotgun sequencing would lead to a tendency
to connect otherwise unrelated graph components; such a bias could be
biological (e.g. repeat present in multiple genomes or other
highly conserved DNA sequences), or non-biological (e.g., inclusion
of sequencing primers in reads or even a low-frequency trend towards
producing specific subsequences \cite{Hansen:2010if,Minoche:2011fl,Dohm:2008ky}).

\begin{figure}[h]
\center
\begin{subfigure}{.5\textwidth}
\centering
\includegraphics[width=\textwidth]{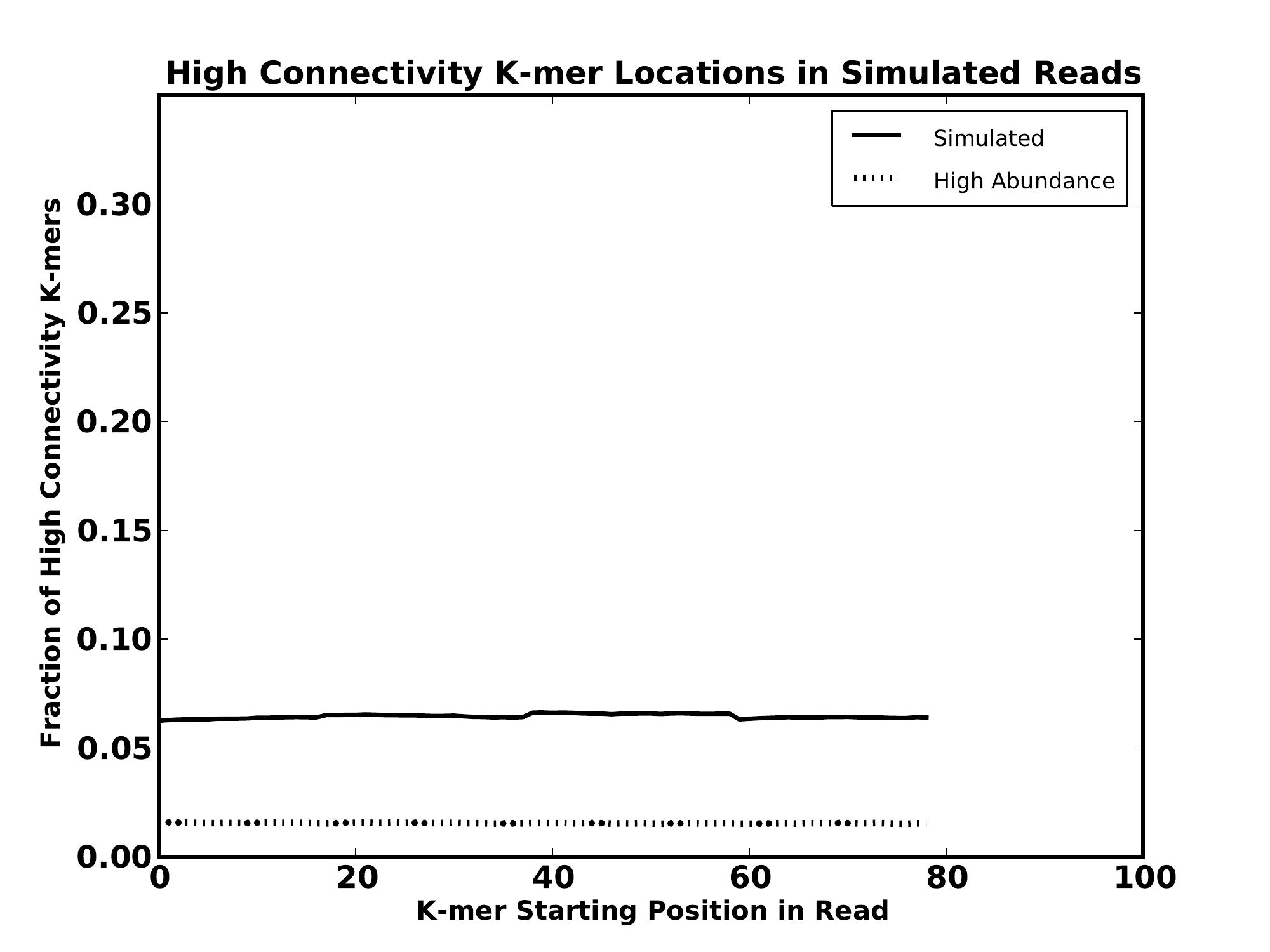}
\end{subfigure}
\begin{subfigure}{.5\textwidth}
\centering
\includegraphics[width=\textwidth]{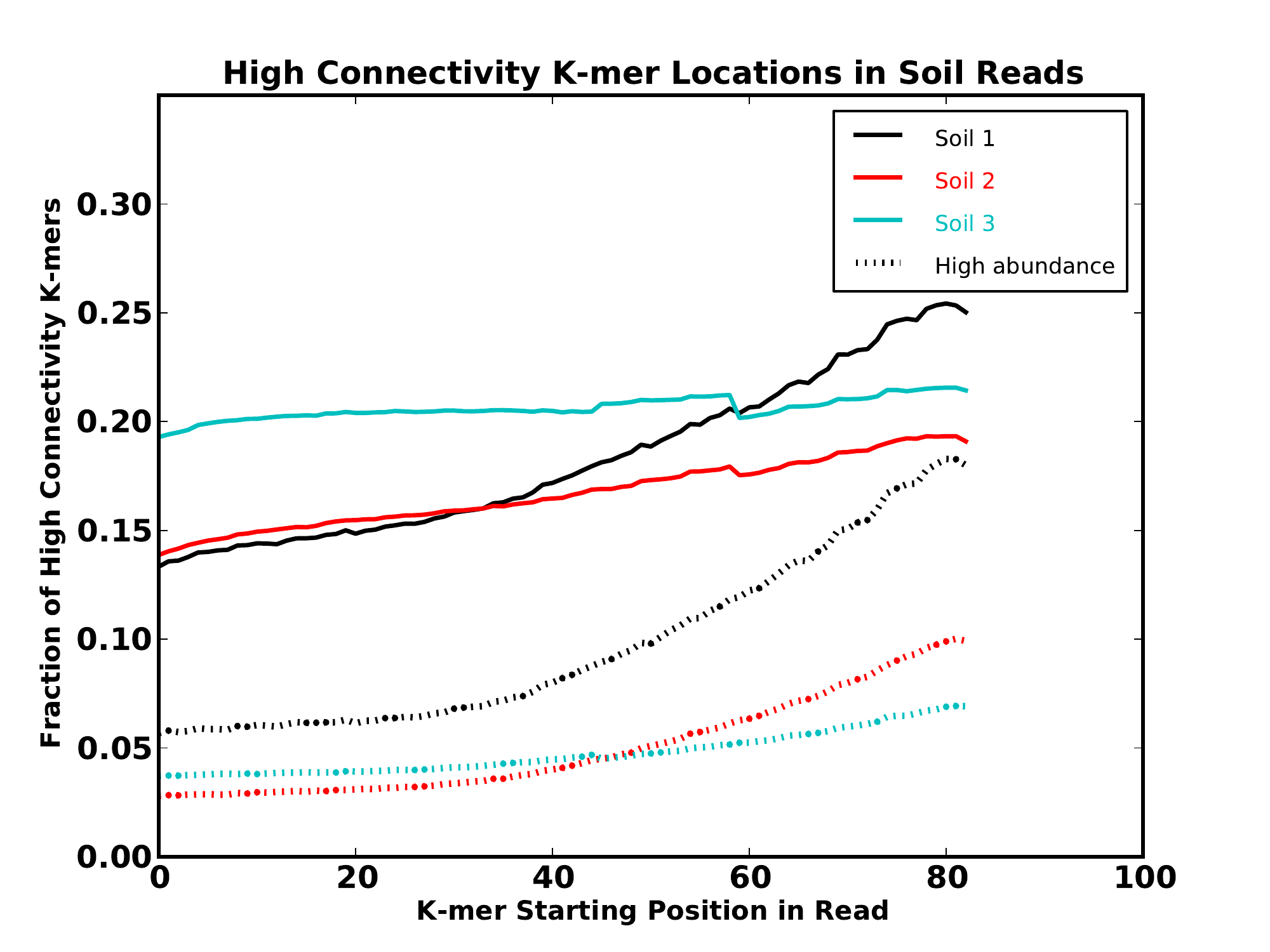}
\end{subfigure}
\begin{subfigure}{.5\textwidth}
\centering
\includegraphics[width=\textwidth]{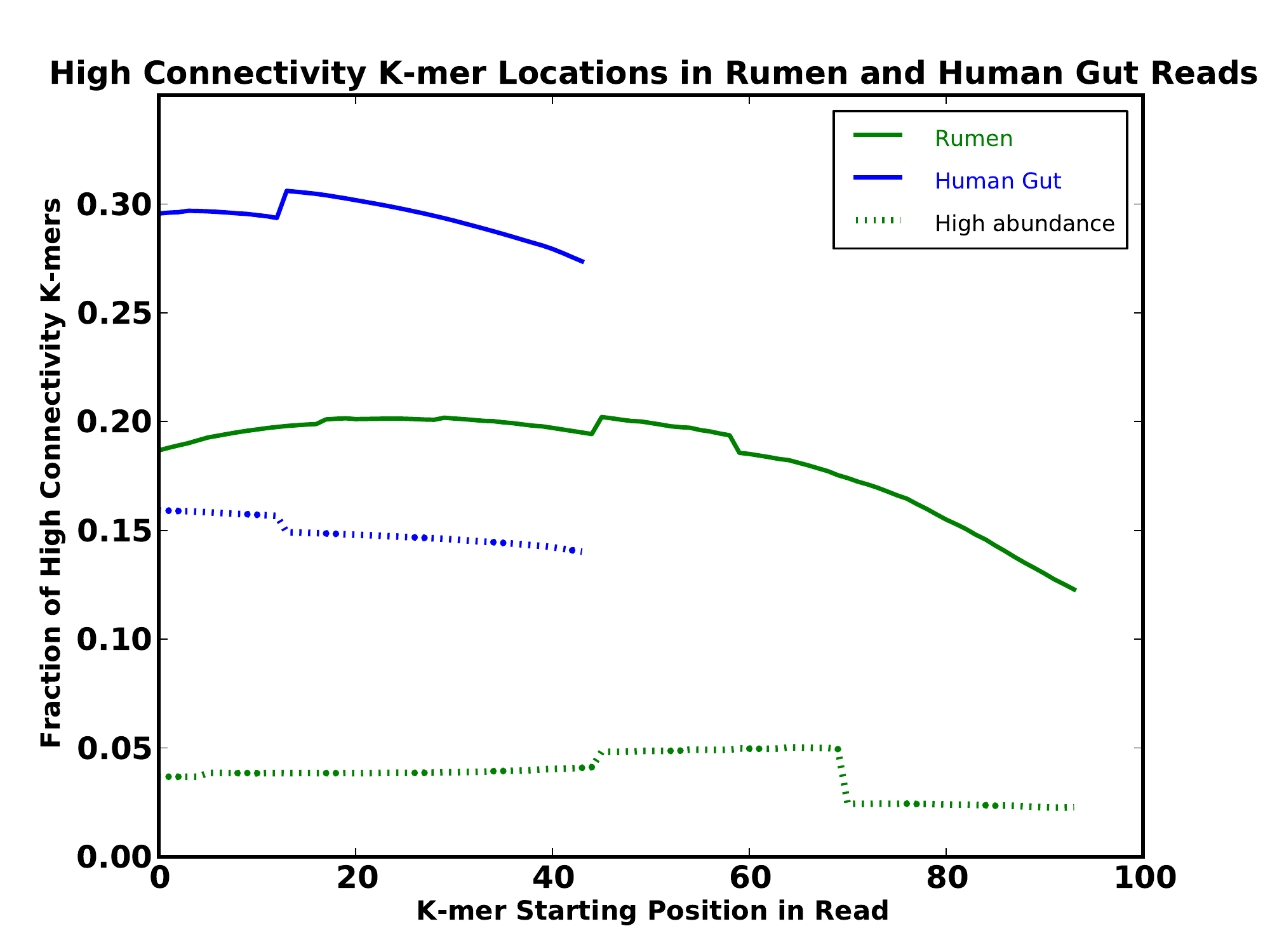}
\end{subfigure}
\caption{The extent to which highly connecting k-mers (solid lines) and the subset of highly abundant (greater than 50) k-mers (dashed lines) are present at specific positions within sequencing reads for various metagenomes.}
\label{pos-spec}
\end{figure}

We believe a significant component of the high connectivity that we
see is of non-biological origin.  Shotgun sequencing is a random
process and consequently any position-specific variation within
sequencing reads is unexpected and probably originates from bias in
sample preparation or the sequencing process \cite{GomezAlvarez:2009p1334, Haas:2011jg, Keegan:2012p1336}.
For the
metagenomes studied here, we used two approaches to examine
characteristics of connectivity correlated to specific positions
within sequencing reads.  First, we measured the connectivity of
sequences at specific positions within reads by calculating local
graph density.  Next, we identified the specific k-mers which were
consistently present in highly dense regions of the assembly graph and
evaluated their location within sequencing reads.  When these
approaches were applied to the simulated dataset, we observed no
position-specific trends when assessing either local graph density
(Fig~\ref{density-pos}) or highly connected k-mers (Fig~\ref{pos-spec}) as is
consistent with the lack of sequencing errors and variation in this
dataset.  In all real metagenomes, however, we identified
position-specific trends in reads for measurements of both local graph density
and the location of highly connected sequences, clearly indicating the
presence of sequencing artifacts.  Although present in all
metagenomes, the direction of the variation varied between soil,
rumen, and human gut datasets, especially for the position-specific
presence of identified highly connected sequences.  It is likely that
there is a larger presence of indirectly preferentially attached reads
which are connected to high coverage sequences of biological origins
in higher coverage datasets, such as the rumen and human gut.  This
preferential attachment of such reads would result in increasing the
number of total reads and consequently the decrease the total fraction
of highly connected k-mers (Fig~\ref{pos-spec}, y-axis).  This trend is observed
in the decreasing fractions of highly connected sequences at the 3'
end of reads as sequencing coverage increased in the small, medium, to
large soil metagenomes and in the soil, rumen, to human gut
metagenomes (Fig~\ref{pos-spec}).

\subsection*{Highly connected sequences are of unknown non-biological origin}

We attempted to identify biological characteristics of highly
connected sequences.  Among the highly connected sequences in the
simulated dataset and those shared by all metagenomes, we identified
only a small fraction (13\% in simulated and less than 7\% in
metagenomes) which matched reference genes associated with core
biological functions (Table~\ref{sim-stoptags} and
~\ref{meta-stoptags}).  This suggests that the remaining sequences are
either not present in known reference genes (i.e., repetitive or
conserved non-coding regions) or originate from non-biological
sources.  This supports the removal of these sequences for typical
assembly and annotation pipelines, where assembly is often followed by
the identification of protein coding regions.

Speculating that many of the highly connected sequences originated
from high abundance reads, we examined the most abundant subsequences.
We found that these subsequences (k-mers present more than 50x in the data set) displayed
similar trends for position-specific variation compared to their
respective sets of highly connected subsequences (Fig~\ref{pos-spec}),
indicating that they contribute significantly to position-specific
variation.  We attempted to identify signatures in these abundant,
highly connected sequences from the simulated and metagenomic datasets by
looking at shorter k-mer profiles.
In the simulated dataset, we found that the total number of unique
5-mers was significantly lower than in metagenomes and that the
most abundant of these 5-mers comprised the large majority of the
total.  This result is consistent with the presence of conserved
biological motifs in the simulated dataset which would result in a
small number of highly abundant sequences; it would also be consistent
with the inclusion of sequencing primers in the data, were this a real
data set.

In contrast, within real metagenomic data, we found that the 5-mers
are evenly distributed and exhibit no specific sequence properties
(Fig~\ref{five-mer}), making them difficult to identify and evaluate.
Most importantly, we were unable to identify any characteristics that
would explain their origin.  In addition, a G-C content analysis of
the highly connective k-mers did not reveal any systematic differences
between the highly connected k-mers and the background k-mer
distribution.

When we reviewed the literature on random and systematic sequencing
errors in Illumina sequencing, we found many different types of
sequencing errors: PCR amplification errors prior to and during
cluster generation; random sequencing errors e.g. from miscalls of
bases; sequencing errors triggered by specific sequence motifs
\cite{Meacham:2011}; adaptor contamination; and post-adaptor read
through.  Of these errors, only random sequencing errors and adaptor
contamination and readthrough would be biased towards the 3' end of
the read.  However, random sequencing error does not contribute to
aberrant de Bruijn graph connectivity \cite{Pell:2012cq}, while
adaptor contamination and readthrough would yield a sharply biased
5-mer distribution.  The observed artifactual sequences thus do not match
any known set of random or systematic errors in Illumina sequencing.

Our current working hypothesis is that a low rate of false connections
are created by a low-frequency tendency towards producing certain
k-mers in the Illumina base calling software, as signal intensities
decline.  We cannot verify this without access to the Illumina
software or source code.

\subsection*{Highly connected sequences are difficult to assemble}

Not all of the observed connectivity within real metagenomes is
artificial, and our approach cannot differentiate between sequencing
artifacts and real biological connectivity.  Therefore, removing
highly connected sequences could remove real biological signal in
addition to sequencing artifacts.  However, we suspected that
assemblers would be unable to generate contigs from highly connected
graph regions, and so even the real sequences would be
underrepresented in the assembly.

Indeed, very few highly connected sequences with abundances greater
than 50 were incorporated into contigs (Table~\ref{assembly-stoptags}). Moreover, those
which were assembled were often disproportionately placed at the ends
of contigs (Fig~\ref{stoptag-contig}), demonstrating that they terminated contig
assembly.  Although this trend was observed for all three assemblers,
it was more prevalent in the Velvet and Meta-IDBA assemblers,
highlighting differences in assembler heuristics.

One initial concern upon discovering this false connectivity was that
these artifacts might nucleate false assemblies, e.g. as seen with the cow
rumen \cite{Hess:2011p686}.  While we only examined their effects on
contig assembly and ignored scaffolding issues, misassembly is an
especially significant concern for environmental metagenomics, where
experimental validation of assemblies is virtually impossible.
However, our observation that k-mers from highly connected graph
regions are generally not present in contigs suggests that these
particular artifacts do not create false assemblies at a high rate.

\subsection*{Filtered assemblies retained most reference genes}

The advantages of removing highly connected sequences must be balanced
against consequences to resulting assemblies.  We compared several
metagenome assemblies before and after the removal of these sequences.
In comparing the simulated dataset's assemblies, the removal of highly
connected sequences resulted in very little loss of annotated
reference genes (less than 1\% total) and some loss of assembled contigs
($\sim$ 15\% of the final assembly).  For the rumen metagenome, we
performed a partial evaluation of the assemblies using available draft
reference genomes.  Similar to the simulated assemblies, we observed
only a small loss (less than 3\% total) of rumen reference genomes assembled
(Table~\ref{assembly-compare}). In general, for all metagenomes, we observed $\sim$ 25\%
loss in assembly after removing highly connected sequences, much more
than observed in assemblies of reference genes and genomes in the
simulated and rumen datasets.  Some of this loss could be beneficial,
resulting from removal of sequencing artifacts.  In addition, trimming
or removing reads from already low-coverage data sets could result in
the loss of contigs due to the length cutoff we use; as observed
above, most of these highly connected k-mers fall at the ends of
contigs, and removing them could shorten the contigs enough to lose
them from our assembly.  It is also possible, of course, that our
approach removes sequences which can accurately be assembled, but we
cannot evaluate this in the absence of reference genomes.

\subsection*{Filtered reads can be assembled more efficiently}

Our original motivation for analyzing connectivity was to assess the
practicality of partitioning metagenomes for later assembly, which can
dramatically reduce the memory requirements for assembly
\cite{Pell:2012cq}.  Not only did removal of highly connected sequences
eliminate problematic sequences, but it resulted in the dissolution
of the largest component and allowed us to partition the metagenomes.

We compared the combined assembly of the partitioned sets of filtered
reads to the original lump dataset, for several assemblers.  For the
partitioned reads, we were able to assemble subsets of reads in
parallel, resulting in significantly reduced time and memory
requirements for assembly (Table~\ref{assembly-stats}).  In the case of the largest soil
metagenome (containing over 500 million reads), we could not complete
the Meta-IDBA assembly of the unfiltered reads in even 100 GB of
memory, but after removing highly connected sequences and
partitioning, the assembly could be completed in less than 2 GB of
memory.  Using partitioned sets of reads for all metagenomes, we were
also able to efficiently complete multiple k-mer length assemblies
(demonstrated with Velvet) and subsequently merge resulting assembled
contigs.  For unfiltered datasets, this was either impossible (due to
memory limitations) or impractical (due to excessive processing time).

\section*{Conclusion}

In this study, we characterize the connectivity of sequences in
several metagenomes to detect and characterize a set of likely
sequencing artifacts.  These artifacts are surprisingly abundant,
comprising 5-13\% of the total unique sequence in real data sets.
Moreover, they have a significant impact on the overall graph
connectivity of the data sets, leading to components containing as
many as 75\% of the reads in the human gut data set.

Despite the prevalence and impact of these sequences, removal of the
highly connected k-mers at the heart of these components leads to
assemblies that are significantly but not catastrophically smaller
than the original assemblies.  The original unfiltered assemblies
contain the majority of the filtered assemblies, while the filtered
assemblies generally contain 70-94\% of the unfiltered assemblies.
The variability in these statistics between the different assemblers
(Table~\ref{assembly-stats}) demonstrates that the assemblers have at least as large an
effect on the content of the assemblies as our filtering procedure.

We cannot reach strong conclusions about the impact of these highly
connected sequences on the correctness of the assembled contigs.  In
particular, in the absence of complex metagenomes that have been
characterized by approaches other than short-read shotgun approaches,
we have no very high quality positive control.  However, we present
evidence that these highly connected sequences primarily affect the
ends of contigs and thus are unlikely to cause incorrect contig
assembly with high frequency.

Our original motivation in exploring metagenome connectivity was to
enable partitioning, an approach that leads to substantially greater
scalability of the assembly procedure.  In this respect, we were successful.
By applying partitioning to filtered metagenome data, we were able to
reduce the maximum memory requirements of assembly (including the
filtering stage) to well below 48 GB of RAM in all cases.  This
enables the use of commodity ``cloud'' computing for all of our
samples \cite{Angiuoli:2011hd}.  The decreased computational
requirements for assembly also enabled ready evaluation of different
assemblers and assembly parameters; as metagenome datasets grow
increasingly larger, this ability to efficiently analyze datasets and
evaluate multiple assemblies will be increasingly important.

More generally, our results demonstrate that metagenome assembly is
still at an early stage of technology development, particularly for
low-coverage data sets.  Different assemblers and different filtering
techniques yield substantially different assembly statistics on these
data sets.  Moreover, there are likely to be additional sources of
sequencing artifacts lurking within large sequencing data sets,
suggesting that more and better computational filtering and validation
approaches need to be developed as environmental metagenomics moves
forward.  Evaluating the assembly graph connectivity created by reads
will be a useful approach in the future.

\section*{Methods}

\subsection*{Metagenomic datasets}
All datasets, with the exception of the agricultural soil metagenome,
originate from previously published datasets. Rumen-associated
sequences (Illumina) were randomly selected from the rumen metagenome (read length 36 - 125 bp)
available at ftp://ftp.jgi-psf.org/pub/rnd2/Cow\_Rumen
\cite{Hess:2011p686}. Human-gut associated sequences (Illumina) of
samples MH0001 through MH0010 were obtained from 
\\*ftp://public.genomics.org.cn/BGI/gutmeta/ Raw\_Reads
\cite{Qin:2010p189} (read length ~44 bp).  The simulated high complexity, high coverage
dataset was previously published \cite{Pignatelli:2011p742}.  Soil metagenomes (read lengths 76-113 bp) are in the SRA (SRX099904 and SRX099905). All
reads used in this study, with the exception of those from the simulated
metagenome, were quality-trimmed for Illumina's read segment quality
control indicator, where a quality score of 2 indicates that all
subsequent regions of the sequence should not be used. After
quality-trimming, only reads with lengths greater than 30 bp were
retained. All quality trimmed datasets, including the previously
unpublished agricultural soil metagenome, are available on a public
Amazon EC2 snapshot (snap-ab88dfdb).  The sequencing coverage of each
metagenome was estimated as the fraction of reads which could be
aligned to assembled contigs with lengths greater than 500 bp.  For
the coverage estimates, an assembly of each metagenome was performed
using Velvet (v1.1.02) with the following parameters: K=33, exp
cov=auto, cov cutoff=0, no scaffolding.  Reads were aligned to
assembled contigs with Bowtie (v0.12.7), allowing for a maximum of two
mismatches.

\subsection*{de Bruijn graph analysis and partitioning software}

We used the probabilistic de Bruijn graph representation previously described by \cite{Pell:2012cq}  to store and partition the metagenome assembly
graphs.  The khmer and screed software packages are required for the analysis,
and the versions used for this publication are available at {\sf https://github.com/ged-lab/khmer/tree/2012-assembly-artifacts} and {\sf https://github.com/ged-lab/screed/tree/2012-assembly-artifacts}.

For metagenomes in this study, we used 4 x 48e9 bit bloom
filters (requiring 24 GB RAM) to store the assembly graphs.  The data processing
pipeline used for this analysis is available for public use on the Amazon Web Services public EBS snapshot snap-ab88dfdb: data-in-paper/lumps and
method-examples/0.partitioning-into-lump.


The local graph density was calculated as the number of
k-mers within a distance of N nodes divided by N. In this
study, N was equal to 10.  For the largest metagenomes, the human gut
and large soil datasets, local graph density was calculated on a
randomly chosen subset of reads because of computational limitations.

To identify specific highly connected sequences within the lump
assembly graphs, graph traversal to a distance of 40 nodes was
attempted from marked waypoints.  If more than 200 k-mers were found
within this traversal were identified (i.e. a graph density $> 5$), all
k-mers within this traversal were marked. If the same k-mers were consistently identified
in other graph traversals, up to five times, the k-mer was flagged as
a highly connected sequence.  Aligning these k-mers to original
sequencing reads, we identified the position-specific location of
these k-mers.  Data and examples of scripts used for this analysis are
available on the Amazon EC2 public snapshot:
data-in-paper/density-bias, data-in-paper/hc-kmer-bias,
method-examples/1.density-analysis,
method-examples/2.identifying-hc-kmers, and
method-examples/3.hc-kmer-analysis.

We identified the sources of highly connected k-mers from the
simulated metagenome by aligning them against reference genes
originating from the 112 source genomes using Bowtie (v0.12.7)
requiring exact matches.  Highly connected k-mers shared between all
the metagenomes were also aligned against the NCBI non-redundant
genome database (ftp://ftp.ncbi.nih.gov/blast/db, March, 1, 2011)
using blastn \cite{Altschul:1990p1335}, requiring an exact match over
the entire k-mer.

We also identified the subset of highly connected 32-mers which were
present greater than 50 times within lumps. Data used for this
analysis are available on the Amazon EC2 public snapshot:
data-in-paper/lumps/HC-kmers/HA-HC-kmers and
method-examples/4.abundant-hc-kmers. These high abundance, highly
connected sequences were aligned to sequencing reads to demonstrate
position specific variation as described above.  We evaluated the
existence of short k-mer (k=5) motifs within high abundance, highly
connected k-mers which did not have an exact match to the NCBI
non-redundant database.  Each identified 32-mer was broken up into
shorter 5-mers, and the frequency of each unique 5-mer was calculated.
Next, each unique 5-mer was ranked based on its abundance, from high
to low, and the cumulative percentage of total 5-mers is shown in the
resulting rank-abundance plot (Fig~\ref{five-mer}).
  
\subsection*{\emph{De novo} metagenomic assembly}

The lump within each dataset was assembled and referred to as the
``unfiltered assembly''.  Additionally, highly connected sequences
identified as described above were were trimmed from sequencing reads
and the remaining reads partitioned and assembled, resulting in the
``filtered assembly''.  \emph{De novo} metagenomic assembly of reads
was completed with Velvet (v1.1.02) with the following parameters:
velveth -short -shortPaired (if applicable to the dataset) and velvetg
-exp\_cov auto -cov\_cutoff 0 -scaffolding no \cite{Zerbino:2008p665}.
For the small and medium soil, rumen, and simulated datasets, Velvet
assemblies were performed at K=25-49, resulting contigs were
dereplicated to remove contigs with 99\% similarity using CD-HIT (v
4.5.6, \cite{Li:2001p1337}), and final contigs were merged with
Minimus (Amos v3.1.0, \cite{Sommer:2007p1253}).  For the largest soil
and human gut metagenomes, assemblies were performed at only K=33 due
to the size of the datasets and memory limitations.  Additional
assemblies were performed with Meta-IDBA (v0.18) \cite{Peng:2011p898}
: --mink 25 --maxk 50 --minCount 0 and with SOAPdenovo: -K 31 -p 8
max\_rd\_len=200 asm\_flags=1 reverse\_seq=0.  After removal of highly
connected k-mers in metagenomic lumps, each filtered lump was
partitioned into separate disconnected subgraphs.  Multiple subgraphs
were grouped together such that assembly could be performed in
parallel on groups of sequences.  Identical assembly parameters and
methods as described above were used for these assemblies.  Unfiltered
and filtered assemblies were compared using the total number of
contigs, total assembly length, and maximum contig size.  Additional,
the coverage of each assembly was calculated through estimating the
average base pair coverage of the BLAST alignment of each assembly to
one another (E-value less than 10$^{-5}$) or, in the case of the
simulated and rumen assemblies, to reference genomes.  The simulated
and rumen reference genomes were previously published in
\cite{Hess:2011p686} and \cite{Pignatelli:2011p742}, respectively.
Resulting assemblies are available on the Amazon EC2 public snapshot:
/data-in-paper/assembly*.

We examined incorporation and the location of the identified high
abundant, highly connecting k-mers within assembled contigs.
Incorporation of these sequences was evaluated by dividing assembled
contigs into words of 32 bp length and identifying exact matches
between sequences and contig fragments.  The location of these k-mers
within assembled unfiltered contigs was examined by dividing each
contig into 100 equally-sized regions.  The fraction of highly
connecting k-mers which aligned exactly to each region was calculated
for each metagenome. Data and examples of scripts used for this
analysis are available on the Amazon EC2 public snapshot:
method-examples/5.hc-kmer-contigs/.

\section*{Acknowledgements}

This project was supported by Agriculture and Food Research Initiative
Competitive Grant no. 2010-65205-20361 from the United States
Department of Agriculture, National Institute of Food and Agriculture
and National Science Foundation IOS-0923812, both to C.T.B.  A.H. was
supported by NSF Postdoctoral Fellowship Award \#0905961 and the Great
Lakes Bioenergy Research Center (Department of Energy BER
DE-FC02-07ER64494).  The work conducted by the U.S. Department of
Energy Joint Genome Institute is supported by the Office of Science of
the U.S. Department of Energy under Contract No. DE-AC02-05CH11231.
We thank Nick Loman and Lex Nederbragt for their helpful comments on
the paper.

\bibliography{artifacts-bib}

\pagebreak

\begin{table}[h]
\centering
\caption{The original size and proportion of highly connective 32-mers in the largest subset of partitioned reads (``lump'') in several medium to high complexity metagenomes.  Read coverage was estimated with the number of aligned sequencing reads to Velvet-assembled contigs (K=33).  The dominant lump, or largest component of each metagenome assembly graph, was found to contain highly connecting (HC) k-mers responsible for high local graph density.  High density nodes refer to nodes with graph density greater than 20.}
\begin{tabular}{l c c c c c c }
& Sm Soil & Med Soil & Large Soil & Rumen & Human Gut & Sim  \\
\hline
Total Reads (millions) & 50.0 & 100.0 & 520.3 & 50.0 & 350.0 & 9.2 \\
Mapped to assembly (percent) & 1.4 & 4.7 & 5.6 & 10.3 & 3.5 & 14.8 \\
Reads in Lump (millions) & 3.0  & 15.0 & 182.2 & 10.3  & 263  & 0.5  \\
Lump Fraction (\%) & 7\% & 15\% & 35\% & 21\% & 75\% & 5\%\\
HC 32-mers (millions) & 6.4 & 33.3 & 230.4 & 25.4 & 136.6 & 0.4\\
Total 32-mers (million) & 84.9 & 326.5 & 2,198.1 & 201.5 & 860.6 & 11.6\\
Fraction of HC 32-mers (\%) & 8\% & 10\% & 10\% & 13\% & 16\% & 3\% \\
High Density Nodes (\%) & 50\% & 37\% & 40\% & 22\% & 28\% & 17\% \\
\hline
\end{tabular}
\label{data-summary}
\end{table}

\begin{table}[h]
\caption{Total number of contigs, assembly length, and maximum contig size was estimated for metagenomic datasets with multiple assemblers, as well as memory and time requirements of unfiltered read assembly (UF).  Filtered reads (F) were processed in 24 GB of memory, and after filtering required less than 2 GB of memory to assemble.  Velvet assemblies of the unfiltered human gut and large soil datasets (marked as *) could only be completed with K=33 due to computational limitations.  The Meta-IDBA assembly of the large soil metagenome could not be completed in less than 100 GB.}
\begin{tabular}{l l l l}
\hline
&UF Assembly &F Assembly 	&UF Requirements \\
& (contigs / length / max size) 	& (contigs / length / max size) 	& Memory (GB)/Time (h)\\
\hline
\emph{Velvet}\\
Small Soil 	&25,470 / 16,269,879 / 118,753	&17,636 / 10,578,908 / 13,246		&5 / 4\\
Medium Soil	&113,613 / 81,660,678 / 57,856	&79,654 / 54,424,264 / 23,663		&18 / 21\\
Large Soil 	&554,825 / 306,899,884 / 41,217 	&290,018 / 159,960,062 /  41,423 	&33 / 12*\\
Rumen		&92,044 / 74,813,072 / 182,003	&72,705 / 49,518,627 / 34,683		&11 / 14\\
Human Gut	&543,331 / 234,686,983  / 85,596	&203,299 / 181,934,800 / 145,740	&76 / 8*\\
Simulated		&11,204 / 6,506,248 / 5,151		&9,859 / 5,463,067 / 6,605		&\textless1 / \textless1\\
\end{tabular}
\medskip
\begin{tabular}{l l l l}
\emph{MetaIDBA} \\ 
Small Soil		&15,739 / 9,133,564 / 37,738 		&12,513 / 7,012,036 / 17,048	&\textless1 / \textless 1 \\
Medium Soil	&76,269 / 45,844,975 / 37,738	 	&52,978 / 30,040,031 / 18,882	&2 / 2\\
Large Soil		&395,122 / 228,857,098 /  37,738	&N/A						&\textgreater116 / incomplete\\
Rumen		&60,330 / 47,984,619 / 54,407		&48,940 / 33,276,502 / 22,083		&12 / 3\\
Human Gut 	&173,432 / 211,067,996 / 106,503	&132,614 / 142,139,101 / 85,539	&58 / 15\\
Simulated		&8,707 / 4,698,575 / 5,113		&7,726 / 4,078,947 / 3,845		&\textless1 / \textless1\\
\end{tabular}
\medskip
\begin{tabular}{l l l l}
\emph{SOAPdenovo}   \\ 
Small Soil		&14,275 / 7,100,052 / 37,720	&12,801 / 6,343,110 / 13,246		&3 / \textless1\\
Medium Soil	&66,640 / 33,321,411 / 28,695	&56,023 / 27,880,293 / 15,721		&10 / \textless1\\
Large Soil		&412,059 / 215,614,765 / 32,514	&334,319 /  171,718,154 / 41,423	&48 / 11\\
Rumen		&62,896 / 40,792,029 / 22,875		&55,975 / 34,540,861 / 19,044	&5 / \textless 1\\
Human Gut	&190,963 / 171,502,574 / 57,803	&161,795 / 139,686,630 / 56,034	&35 / 5\\
Simulated		&6,322 / 2,940,509 / 3,786		&6,029 / 2,821,631 / 3,764	&\textless1 / \textless1\\
\end{tabular}
\label{assembly-stats}
\end{table}

\begin{table}[h]
\caption{Comparison of unfiltered (UF) and filtered (F) assemblies of various metagenome lumps using Velvet, SOAPdenovo, and Meta-IDBA assemblers.  Assemblies were aligned to each other, and coverage was estimated (columns 1-2).  Simulated and rumen assemblies were aligned to available reference genomes (RG) (columns 3-4).}
Velvet Assembler  \\
\begin{tabular}{l c c c c}
\hline
& Cov. of UF by F & Cov. of F by UF  & Cov. of RG  by UF  & Cov. of RG by F \\
\hline
Simulated		&85.4\%		&99.4\%		&5.4\%	&4.6\%\\
Small Soil		&74.7\%		&98.8\%		&-		&-\\
Medium Soil	&75.6\%		&98.4\%		&-		&-\\
Large Soil		&50.9\%		&86.6\%		&-		&-\\
Rumen		&75.9\%		&98.8\%		&17.5\%	&14.8\%\\
Human Gut 	&80.0\%		&89.1\%		&-		&-\\
\end{tabular}
\medskip
Meta-IDBA Assembler  \\
\begin{tabular}{l c c c c}
\hline
& Cov. of UF by F & Cov. of F by UF  & Cov. of RG  by UF  & Cov. of RF by F \\
\hline
Simulated		&87.4\%		&94.4\%		&4.7\%	&4.1\%\\
Small Soil		&75.7\%		&94.2\%		&-		&-\\
Medium Soil	&67.7\%		&94.8\%		&-		&-\\
Large Soil		&N/A			&N/A			&-		&-\\
Rumen		&70.8\%		&95.0\%		&17.5\%	&14.8\%\\
Human Gut 	&74.4\%		&99.4\%		&-		&-\\
\end{tabular}
\medskip
SOAPdenovo Assembler \\   
\begin{tabular}{l c c c c}
\hline
& Cov. of UF by F & Cov. of F by UF  & Cov. of RG  by UF  & Cov. of RF by F \\
\hline
Simulated		&94.0\%		&97.0\%		&3.0\%	&2.9\%\\
Small Soil		&86.8\%		&96.1\%		&-		&-\\
Medium Soil	&82.4\%		&96.0\%		&-		&-\\
Large Soil		&78.9\%		&94.5\%		&-		&-\\
Rumen		&85.2\%		&97.8\%		&14.9\%	&13.6\%\\
Human Gut 	&85.4\%		&99.3\%		&-		&-\\
\end{tabular}
\label{assembly-compare}
\end{table}

\begin{table}[h]
\caption{Total number of abundant (greater than 50x) highly connective sequences incorporated into unfiltered assemblies}
\begin{tabular}{l c c c}
 & Velvet & SOAPdenovo & MetaIDBA  \\
\hline
Small Soil & 0 (0.0\%) & 41 (0.0\%) & 8,717 (0.1\%) \\
Medium Soil &  32,328 (0.1\%) & 852 (0.0\%) & 23,881 (0.1\%) \\
Large Soil & 643,071 (0.3\%) & 279,519 (0.1\%) & N/A \\
Rumen & 45,721 (0.2\%) & 14,858 (0.1\%) & 33,046 (0.1\%) \\
Human Gut & 4,661,447 (3.4\%) & 1,749,347 (1.3\%) & 5,528,054 (4.0\%) \\
Simulated & 5,118 (1.4\%) & 0 (0.0\%) & 5,480 (1.5\%) \\
\hline
\end{tabular}
\label{assembly-stoptags}
\end{table}

\begin{table}
\caption{Annotations (against 112 reference genomes) of highly-connecting (HC) sequences identified in the simulated metagenome.}
\begin{tabular}{l c}
\hline
& Number of HC sequences with annotation\\
\hline
ABC transporter-like protein	&306\\
Methyl-accepting chemotaxis sensory transducer	&210\\
ABC transporter	&173\\
Elongation factor Tu	&94\\
Chemotaxis sensory transducer	&51\\
ABC transporter ATP-binding protein	&44\\
Diguanylate cyclase/phosphodiesterase	&36\\
ATPase	&36\\
S-adenosyl-L-homocysteine hydrolase	&36\\
Adenosylhomocysteine And downstream NAD binding	&36\\
Ketol-acid reductoisomerase	&34\\
S-adenosylmethionine synthetase	&34\\
Elongation factor G	&34\\
ABC transporter ATPase	&33\\
\end{tabular}
\label{sim-stoptags}
\end{table}

\begin{table}
\caption{Annotations (against NCBI-nr database) of highly-connecting (HC) sequences identified in thee three soil, rumen, and human gut metagenomes.}
\begin{tabular}{l c}
\hline
& Number of HC sequences with annotation \\
\hline
Translation elongation factor/GTP-binding protein LepA	&11\\
S-adenosylmethionine synthetase	&8\\
Aspartyl-tRNA synthetase	 &8\\
Malate dehydrogenase	&7\\
V-type H(+)-translocating pyrophosphatase	&6\\
Acyl-CoA synthetase	&6\\
NAD synthetase 	&5\\
Ribonucleotide reductase of class II	&4\\
Ribityllumazine synthase	&4\\
Heavy metal translocating P-type ATPase, copA	&3\\
GyrB	 &3\\
Glutamine amidotransferase chain of NAD synthetase	&3\\
ChaC family protein	&3\\
\end{tabular}
\label{meta-stoptags}
\end{table}

\end{document}